\documentclass[conference]{IEEEtran}
\IEEEoverridecommandlockouts
\usepackage{cite}
\usepackage{amsmath,amssymb,amsfonts}
\usepackage{algorithmic}
\usepackage{graphicx}
\usepackage{textcomp}
\usepackage{xcolor}
\usepackage{url}
\usepackage{hyperref}

\def\BibTeX{{\rm B\kern-.05em{\sc i\kern-.025em b}\kern-.08em
    T\kern-.1667em\lower.7ex\hbox{E}\kern-.125emX}}

\makeatletter
\newcommand{\linebreakand}{%
  \end{@IEEEauthorhalign}
  \hfill\mbox{}\par
  \mbox{}\hfill\begin{@IEEEauthorhalign}
}
\makeatother

\begin{document}

\title{ShikkhaChain: A Blockchain-Powered Academic Credential Verification System for Bangladesh}
\author{
\IEEEauthorblockN{Ahsan Farabi}
\IEEEauthorblockA{\textit{Dept. of CSE} \\
\textit{United International University} \\
Dhaka, Bangladesh \\
afarabi221266@bscse.uiu.ac.bd}
\and
\IEEEauthorblockN{Israt Khandaker}
\IEEEauthorblockA{\textit{Dept. of CSE} \\
\textit{United International University} \\
Dhaka, Bangladesh \\
ikhandaker221263@bscse.uiu.ac.bd}
\and
\IEEEauthorblockN{Jayed Ahsan}
\IEEEauthorblockA{\textit{Dept. of CSE} \\
\textit{United International University} \\
Dhaka, Bangladesh \\
jahsan221125@bscse.uiu.ac.bd}
\linebreakand
\IEEEauthorblockN{Ibrahim Khalil Shanto}
\IEEEauthorblockA{\textit{Dept. of CSE} \\
\textit{United International University} \\
Dhaka, Bangladesh \\
ishanto213193@bscse.uiu.ac.bd}
\and
\IEEEauthorblockN{Nusrat Jahan}
\IEEEauthorblockA{\textit{Dept. of CSE} \\
\textit{United International University} \\
Dhaka, Bangladesh \\
njahan221323@bscse.uiu.ac.bd}
\and
\IEEEauthorblockN{Md. Jarif Khan}
\IEEEauthorblockA{\textit{Dept. of CSE} \\
\textit{BRAC University} \\
Dhaka, Bangladesh \\
md.jarif.khan@g.bracu.ac.bd}
}

\maketitle

\begin{abstract}
Academic credential fraud threatens educational integrity, especially in developing countries like Bangladesh, where verification methods are primarily manual and inefficient. To address this challenge, we present \textbf{ShikkhaChain}—a blockchain-powered certificate management platform designed to securely issue, verify, and revoke academic credentials in a decentralized and tamper-proof manner. Built on Ethereum smart contracts and utilizing IPFS for off-chain storage, the platform offers a transparent, scalable solution accessible through a React-based DApp with MetaMask integration. ShikkhaChain enables role-based access for governments, regulators, institutions, and public verifiers, allowing QR-based validation and on-chain revocation tracking. Our prototype demonstrates enhanced trust, reduced verification time, and improved international credibility for Bangladeshi degrees, promoting a more reliable academic and employment ecosystem.
\end{abstract}


\section{Introduction}
Verifying academic credentials is a cornerstone of trust in education and employment systems. In Bangladesh, however, verification remains dominated by manual and paper-based processes that are slow, inefficient, and vulnerable to forgery. The absence of robust digital infrastructure has led to credential fraud, delays in recruitment, and reduced international credibility of Bangladeshi degrees. A scalable, tamper-resistant mechanism for issuing and validating certificates is therefore essential to strengthen both domestic and global trust in the education system.

Recent studies have explored blockchain-based approaches to credential management, demonstrating benefits such as immutability, decentralization, and transparency \cite{rahman2023verifi, park2023blockcerts, habib2024credsec}. Yet significant gaps persist: existing systems often lack role-aware governance (e.g., distinguishing governments, regulators, institutions, and public verifiers), overlook integration with decentralized storage like IPFS for scalability, and rarely align with localized regulatory and infrastructural needs. These limitations reduce their applicability in national contexts such as Bangladesh.

To address these challenges, we present \textbf{ShikkhaChain}, a blockchain-powered credential verification system tailored for Bangladesh. ShikkhaChain leverages Ethereum smart contracts to issue, verify, and revoke certificates, while IPFS provides decentralized off-chain storage linked by on-chain content identifiers (CIDs). A React-based decentralized application (DApp) with MetaMask integration offers secure, role-based access for government authorities, regulators, institutions, graduates, and employers. The system ensures transparent, tamper-proof, and efficient verification, promoting greater trust in both academic and employment ecosystems.

The main contributions of this paper are as follows:
\begin{itemize}
    \item Design of a role-aware, layered architecture that integrates government, regulator, institutional, and public roles for transparent credential governance.
    \item Development of Ethereum smart contracts that support certificate issuance, verification, and revocation with event-driven off-chain indexing.
    \item Integration of IPFS for scalable and tamper-resistant off-chain storage linked to on-chain records.
    \item Implementation of a functional prototype featuring a React-based DApp with QR code and hash-based verification.
    \item Comparative analysis with prior blockchain-based systems, highlighting improvements in role-awareness, revocation, and transparency.
\end{itemize}

The remainder of this paper is structured as follows: Section II reviews related work on blockchain credential systems. Section III presents the system architecture and design. Section IV details the prototype results and comparative analysis. Section V discusses security and trust features. Section VI outlines limitations and future directions. Section VII concludes the paper.

\section{Related Work}
The application of blockchain in higher education has gained momentum due to its potential for data integrity, decentralized trust, and streamlined credential verification. We group prior work into: (i) credential verification systems, (ii) decentralized frameworks and IPFS integration, (iii) governance and access control, and (iv) smart-contract implementations.

\subsection{Credential Verification and Academic Trust}
Blockchain has been widely adopted to establish trust and prevent tampering in educational credentials. Centeno and Palaoag \cite{centeno2024blockchain} designed a prototype for blockchain-based degree verification. Xu \cite{xu2024development} proposed a decentralized credential authentication framework to reduce fraud. Shivarkar \cite{shivarkar2025academic} proposed a decentralized digital certification system replacing paper-based certificates. Nguyen and Lin \cite{nguyen2023eduledger} developed EduLedger for decentralized transcripts. These solutions demonstrate security benefits but often overlook stakeholder role differentiation and modular policy integration.

\subsection{Systematic Reviews and Global Implementations}
Silaghi and Popescu \cite{silaghi2025systematic} reviewed blockchain initiatives in higher education, highlighting scalability and regulatory adoption gaps. De Alwis et al. \cite{de2025exploring} emphasized governance for accreditation. Country-specific case studies include a UAE national platform \cite{al2024blockchain} and BlockMEDC for Moroccan institutions \cite{fartitchou2024blockmedc}, which show feasibility yet limited generalizability and standardized role-based governance.

\subsection{Decentralization, IPFS, and Data Provenance}
Decentralized storage like IPFS mitigates centralized failures. Rahman et al. \cite{rahman2023verifi} combined Ethereum and IPFS for tamper-resistant records; Ahmad and Lee \cite{ahmad2024trust} proposed TrustChainEdu (IPFS–blockchain hybrid). Nasir and Bukhari \cite{nasir2024ipfs} linked IPFS with on-chain proof for provenance. These works evolve secure architectures but do not address policy-level differentiation between ministries, universities, and employers—addressed in our design.

\subsection{Smart Contracts, Revocation, and Role-Based Access}
Smart contracts enable automated issuance and revocation. Park and Zhang \cite{park2023blockcerts} introduced BlockCerts+ with revocation; Mohapatra and Sen \cite{mohapatra2024smart} designed SmartCertEdu embedding issuance, storage, and verification with IPFS; Habib et al. \cite{habib2024credsec} proposed CredSec for stakeholder-friendly access; Flanery et al. \cite{flanery2023web} framed learner-centered Web3 credential control. Gaps remain in integrating role-aware contracts, regulator-controlled institution lists, and publicly verifiable yet privacy-preserving records—gaps addressed by ShikkhaChain.

\section{System Architecture and Design}
\begin{figure}[t]
    \centering
    \includegraphics[width=0.9\linewidth]{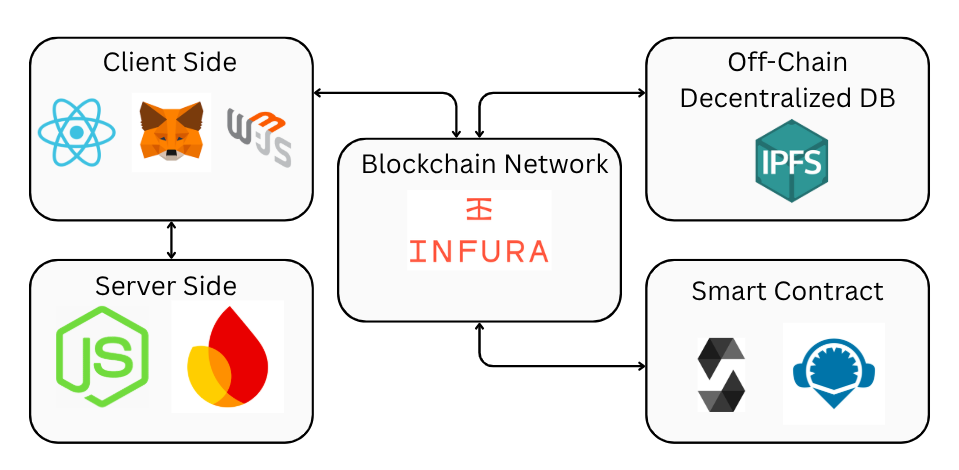}
    \caption{Layered system architecture of ShikkhaChain, illustrating stakeholder roles, blockchain/IPFS integration, and modular access layers.}
    \label{fig:architecture}
\end{figure}

\begin{figure}[t]
    \centering
    \includegraphics[width=0.9\linewidth]{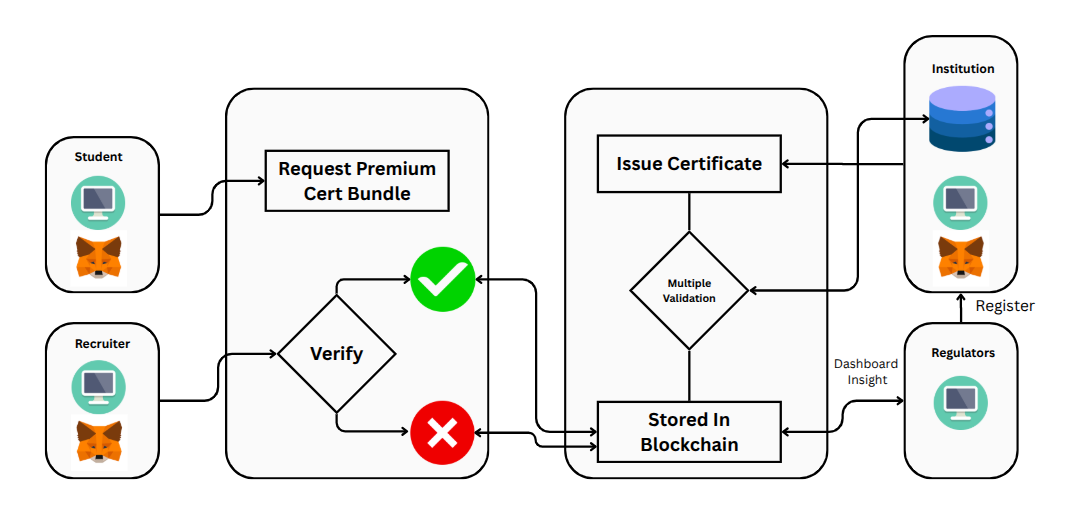}
    \caption{Workflow of certificate issuance and verification in ShikkhaChain. Institutions issue credentials, which are stored on IPFS and linked on-chain, while verifiers confirm authenticity through the DApp.}
    \label{fig:flow_chart}
\end{figure}

ShikkhaChain is designed to issue, verify, and revoke academic certificates using blockchain and Web3 technologies. The architecture is modular, layered, and role-based, ensuring scalability, policy enforcement, and decentralized trust (Fig.~\ref{fig:architecture}). In addition, the workflow for issuance and verification is illustrated in Fig.~\ref{fig:flow_chart}.

\subsection{Stakeholder Roles}
The system defines distinct roles to establish governance, accountability, and role-based access:
\begin{itemize}
    \item \textbf{Government:} Acts as the root authority, responsible for onboarding regulators, setting national policies, and overseeing compliance across the ecosystem.
    \item \textbf{Regulators:} Serve as intermediaries, maintaining an up-to-date registry of authorized institutions, enforcing issuance policies, and auditing institutional activities.
    \item \textbf{Institutions:} Universities and colleges authorized by regulators can issue new certificates or revoke fraudulent/invalid ones. They interact directly with the smart contracts.
    \item \textbf{Public (Graduates/Employers):} Graduates present verifiable digital certificates to employers, while employers or any verifier can check authenticity through the DApp using a QR code or hash.
\end{itemize}
This stakeholder model ensures hierarchical trust delegation, where control flows from the government down to the end-users without a central database.

\subsection{System Layers}
The layered architecture separates responsibilities to achieve modularity and extensibility:
\begin{enumerate}
    \item \textbf{User Layer:} Provides browser-based interaction through a React.js interface, with MetaMask wallet integration for authentication and transaction signing.
    \item \textbf{Application Layer:} Implements dashboards for different roles (government, regulator, institution, public). It manages certificate requests, verification results, and revocation operations, ensuring a user-friendly interface.
    \item \textbf{Access Layer:} Handles blockchain connectivity using Web3.js and ethers.js. This layer supports communication with Ethereum nodes via providers such as Infura or Alchemy and ensures secure MetaMask wallet connectivity.
    \item \textbf{Blockchain Layer:} Contains the Ethereum smart contracts that encode core logic for certificate issuance, verification, and revocation. State changes (e.g., a revoked certificate) are immutable and transparent.
    \item \textbf{Off-chain Storage Layer:} Stores certificate metadata (e.g., graduate name, institution, degree type) on IPFS. The content identifier (CID) is stored on-chain, providing both scalability and tamper-resistance.
\end{enumerate}
This separation of concerns improves maintainability and supports future upgrades such as Layer-2 migration.

\subsection{Smart Contract Design}
ShikkhaChain’s smart contracts implement fine-grained access control and certificate lifecycle management:
\begin{itemize}
    \item \texttt{issueCertificate()}: Called by an authorized institution to issue a certificate. The metadata CID is recorded on-chain, and an event is emitted for off-chain indexing.
    \item \texttt{verifyCertificate()}: Publicly accessible function to validate whether a certificate hash corresponds to a registered CID and has not been revoked.
    \item \texttt{revokeCertificate()}: Invoked by authorized institutions to mark a certificate as invalid. The status change is recorded immutably on-chain.
\end{itemize}
Contracts enforce role checks at runtime, ensuring only permitted entities can perform sensitive actions. Additionally, emitted events (e.g., \texttt{CertificateIssued}, \texttt{CertificateRevoked}) enable external services to build audit trails and analytics.

\subsection{Verification Process}
The verification workflow is designed to be lightweight and user-friendly:
\begin{enumerate}
    \item The verifier scans the certificate QR code or inputs the certificate hash into the DApp.
    \item The DApp retrieves the associated metadata CID from the blockchain.
    \item Using the CID, the DApp fetches the certificate metadata stored on IPFS.
    \item The smart contract is queried to check certificate status (valid, revoked, or not found).
    \item The system displays verification results, including certificate details, issuer, and current status.
    \item Optionally, the verified result can be exported as a digitally signed PDF using jsPDF, suitable for archiving or submission to external parties.
\end{enumerate}
This process eliminates manual verification delays and ensures real-time authenticity checks without dependence on centralized authorities.

\section{Results and Comparative Analysis}
We implemented a prototype of \textbf{ShikkhaChain} using React.js for the decentralized application (DApp), Solidity smart contracts on the Ethereum Goerli test network, and IPFS for off-chain storage. MetaMask was used for secure wallet-based authentication and transaction signing.

\subsection{Functional Outcomes}
The prototype successfully demonstrated the complete lifecycle of academic certificates:
\begin{itemize}
    \item \textbf{Issuance:} Authorized institutions upload certificate metadata to IPFS. The generated content identifier (CID) is immutably stored on-chain through a smart contract.
    \item \textbf{Verification:} Employers or public users validate certificates by entering a hash or scanning a QR code. The system fetches the CID from the blockchain, retrieves metadata from IPFS, and confirms the certificate’s validity.
    \item \textbf{Revocation:} Institutions revoke compromised or invalid certificates through \texttt{revokeCertificate()}, and the updated status is recorded immutably on-chain.
    \item \textbf{Export:} Verified results can be exported as digitally signed PDFs for archiving or external submission.
\end{itemize}

\subsection{Comparative Feature Analysis}
Table~\ref{tab:comparison} highlights the differences between ShikkhaChain and representative blockchain-based credential systems. While prior systems provide decentralized storage and verification, they typically lack role differentiation and on-chain revocation tracking. ShikkhaChain uniquely integrates:
\begin{enumerate}
    \item \textbf{Role-aware access control} across four distinct stakeholders (government, regulator, institution, public).
    \item \textbf{IPFS-backed storage} with on-chain CIDs for scalability and tamper-resistance.
    \item \textbf{On-chain revocation} that ensures transparent invalidation of compromised certificates.
\end{enumerate}

\begin{table}[t]
\centering
\caption{Comparison of blockchain-based credential systems}
\label{tab:comparison}
\begin{tabular}{@{}l l l@{}}
\hline
\textbf{System} & \textbf{Storage} & \textbf{Role Access} \\
\hline
\textbf{ShikkhaChain} & IPFS + on-chain CID & 4 Major Roles\\
Verifi-Chain \cite{rahman2023verifi} & IPFS & No roles \\
BlockCerts+ \cite{park2023blockcerts} & On-chain hash + local file & Issuer-only \\
CredSec \cite{habib2024credsec} & IPFS & Admin, Issuer, Verifier \\
BlockMEDC \cite{fartitchou2024blockmedc} & IPFS & No roles \\
\hline
\end{tabular}
\end{table}

\subsection{Performance Insights}
Performance evaluation on the Goerli Ethereum test network showed the following:
\begin{itemize}
    \item \textbf{Transaction Time:} 12--25 seconds per transaction, depending on network congestion.
    \item \textbf{Gas Cost:} Approximately \(\sim 1{,}289{,}600\) gas units to issue a certificate. On Ethereum mainnet, this would incur significant cost, motivating adoption of Layer-2 solutions.
    \item \textbf{IPFS Latency:} 1--5 seconds for metadata retrieval, influenced by pinning and gateway reliability.
\end{itemize}

These results confirm the feasibility of ShikkhaChain, while also indicating that scaling to production environments would benefit from optimizations such as Layer-2 deployment, dedicated IPFS pinning, or a permissioned blockchain for national-level adoption.

\section{Security and Trust Features}
\begin{itemize}
    \item \textbf{Immutability:} Blockchain anchoring ensures tamper-resistance.
    \item \textbf{Content Integrity:} IPFS CIDs guarantee file integrity.
    \item \textbf{Transparency:} Public verification without intermediaries.
    \item \textbf{Decentralized Access:} Wallet-based authentication; no central DB of secrets.
\end{itemize}

\section{Discussion and Limitations}
Although the prototype demonstrates the feasibility of ShikkhaChain, several limitations remain that must be addressed for production-scale adoption:
\begin{itemize}
    \item \textbf{On-chain cost and latency:} Transactions on the Ethereum mainnet incur high gas costs and confirmation delays. While acceptable for a prototype, large-scale deployment would require migration to Layer-2 solutions or permissioned blockchains.
    \item \textbf{IPFS availability:} Certificate metadata retrieval depends on IPFS pinning and gateway reliability, which may cause delays or temporary inaccessibility in practice.
    \item \textbf{Governance bootstrapping:} The hierarchical model requires effective onboarding of regulators and institutions. Without proper adoption and compliance enforcement, the trust model may weaken.
    \item \textbf{Privacy concerns:} Current design exposes certificate metadata publicly once the CID is known. Privacy-preserving verification (e.g., via zk-SNARKs) is needed to protect sensitive details while still proving authenticity.
    \item \textbf{Integration challenges:} ShikkhaChain currently functions as a standalone prototype. Standardized APIs and interoperability with national e-Governance systems would be necessary for real-world deployment.
\end{itemize}

\section{Conclusion and Future Work}
This paper presented \textbf{ShikkhaChain}, a blockchain-powered credential verification system for Bangladesh. By combining Ethereum smart contracts, IPFS storage, role-aware access, and QR-enabled verification, the system ensures secure issuance, transparent validation, and immutable revocation of academic certificates. Comparative analysis shows that ShikkhaChain advances prior solutions through multi-tiered stakeholder roles and on-chain revocation tracking.

Future work will focus on: (i) scalability via Layer-2 or permissioned networks, (ii) privacy with zero-knowledge proofs, (iii) mobile-first DApp with offline QR support, (iv) interoperability with national databases, (v) AI-driven anomaly detection to identify fraudulent patterns, and (vi) formal verification and third-party audits of smart contracts. These directions will help ShikkhaChain mature into a production-ready platform capable of securing academic integrity at national scale.

\section*{Data and Code Availability}
All source code, smart contract artifacts, and demonstration materials for ShikkhaChain are publicly available at:  
\url{https://github.com/TheAhsanFarabi/ShikkhaChain}

\bibliographystyle{IEEEtran}
\bibliography{references}

\begin{thebibliography}{10}
\providecommand{\url}[1]{#1}
\csname url@samestyle\endcsname
\providecommand{\newblock}{\relax}
\providecommand{\bibinfo}[2]{#2}
\providecommand{\BIBentrySTDinterwordspacing}{\spaceskip=0pt\relax}
\providecommand{\BIBentryALTinterwordstretchfactor}{4}
\providecommand{\BIBentryALTinterwordspacing}{\spaceskip=\fontdimen2\font plus
\BIBentryALTinterwordstretchfactor\fontdimen3\font minus \fontdimen4\font\relax}
\providecommand{\BIBforeignlanguage}[2]{{%
\expandafter\ifx\csname l@#1\endcsname\relax
\typeout{** WARNING: IEEEtran.bst: No hyphenation pattern has been}%
\typeout{** loaded for the language `#1'. Using the pattern for}%
\typeout{** the default language instead.}%
\else
\language=\csname l@#1\endcsname
\fi
#2}}
\providecommand{\BIBdecl}{\relax}
\BIBdecl

\bibitem{rahman2023verifi}
\BIBentryALTinterwordspacing
T.~Rahman, S.~I. Mouno, A.~M. Raatul, A.~K. Al~Azad, and N.~Mansoor, ``Verifi-chain: A credentials verifier using blockchain and ipfs,'' \emph{arXiv preprint arXiv:2307.05797}, 2023. [Online]. Available: \url{https://arxiv.org/abs/2307.05797}
\BIBentrySTDinterwordspacing

\bibitem{park2023blockcerts}
\BIBentryALTinterwordspacing
M.~Park and L.~Zhang, ``Blockcerts+: A secure blockchain-based certificate system with smart contract revocation,'' \emph{IEEE Access}, vol.~11, pp. 134\,756--134\,770, 2023. [Online]. Available: \url{https://ieeexplore.ieee.org/document/10172612}
\BIBentrySTDinterwordspacing

\bibitem{habib2024credsec}
\BIBentryALTinterwordspacing
M.~A. Habib, M.~M. Rahman, and N.~H. Neom, ``Credsec: A blockchain-based secure credential management system for university adoption,'' \emph{arXiv preprint arXiv:2406.05151}, 2024. [Online]. Available: \url{https://arxiv.org/abs/2406.05151}
\BIBentrySTDinterwordspacing

\bibitem{centeno2024blockchain}
\BIBentryALTinterwordspacing
K.~Centeno~Cuya and T.~D. Palaoag, ``Blockchain in higher education: Advancing security, verification, and trust in academic credentials,'' \emph{Nanotechnology Perceptions}, 2024, published: May 2024. [Online]. Available: \url{https://www.researchgate.net/publication/389949495_Blockchain_ensuring_academic_integrity_with_a_degree_verification_prototype}
\BIBentrySTDinterwordspacing

\bibitem{xu2024development}
\BIBentryALTinterwordspacing
Y.~Xu, ``Development of blockchain-based academic credential verification system,'' \emph{Open Access Library Journal}, vol.~11, no.~09, pp. 1--20, 2024. [Online]. Available: \url{https://www.researchgate.net/publication/384476272_Development_of_Blockchain-Based_Academic_Credential_Verification_System}
\BIBentrySTDinterwordspacing

\bibitem{shivarkar2025academic}
\BIBentryALTinterwordspacing
S.~Shivarkar, ``Academic certificate verification using decentralized digital certification,'' \emph{International Journal of Scientific Research in Engineering and Management}, vol.~9, no.~2, pp. 1--9, 2025. [Online]. Available: \url{https://www.researchgate.net/publication/389229234_Academic_Certificate_Verification_Using_Decentralized_Digital_Certification}
\BIBentrySTDinterwordspacing

\bibitem{nguyen2023eduledger}
\BIBentryALTinterwordspacing
T.~Nguyen and X.~Lin, ``Eduledger: Blockchain-based decentralized transcript management system,'' \emph{Information Sciences}, vol. 644, pp. 119--138, 2023. [Online]. Available: \url{https://doi.org/10.1016/j.ins.2023.04.018}
\BIBentrySTDinterwordspacing

\bibitem{silaghi2025systematic}
\BIBentryALTinterwordspacing
D.~L. Silaghi and D.~E. Popescu, ``A systematic review of blockchain-based initiatives in comparison to best practices used in higher education institutions,'' \emph{Computers}, vol.~14, no.~4, p. 141, 2025. [Online]. Available: \url{https://www.mdpi.com/2073-431X/14/4/141}
\BIBentrySTDinterwordspacing

\bibitem{de2025exploring}
\BIBentryALTinterwordspacing
A.~de~Alwis, A.~Shrestha, and T.~Sarker, ``Exploring governance for accreditation in the education sector using blockchain technology: A systematic literature review,'' \emph{Discover Education}, vol.~4, no.~57, 2025. [Online]. Available: \url{https://link.springer.com/article/10.1007/s44217-025-00449-y}
\BIBentrySTDinterwordspacing

\bibitem{al2024blockchain}
\BIBentryALTinterwordspacing
M.~Al~Hemairy, M.~Abu~Talib, A.~Khalil, A.~Zulfiqar, and T.~Mohamed, ``Blockchain-based framework and platform for validation, authentication \& equivalency of academic certification and institution's accreditation: Uae case study and system performance,'' \emph{Education and Information Technologies}, vol.~29, pp. 18\,203--18\,232, 2024. [Online]. Available: \url{https://link.springer.com/article/10.1007/s10639-024-12493-6}
\BIBentrySTDinterwordspacing

\bibitem{fartitchou2024blockmedc}
\BIBentryALTinterwordspacing
M.~Fartitchou, I.~Lamaakal, K.~El~Makkaoui, Z.~El~Allali, and Y.~Maleh, ``Blockmedc: Blockchain smart contracts for securing moroccan higher education digital certificates,'' \emph{arXiv preprint arXiv:2410.07258}, 2024. [Online]. Available: \url{https://arxiv.org/abs/2410.07258}
\BIBentrySTDinterwordspacing

\bibitem{ahmad2024trust}
\BIBentryALTinterwordspacing
F.~Ahmad and H.~Lee, ``Trustchainedu: Decentralized trust model for educational credential verification using blockchain and ipfs,'' \emph{Computers and Education: Artificial Intelligence}, vol.~5, p. 100112, 2024. [Online]. Available: \url{https://doi.org/10.1016/j.caeai.2024.100112}
\BIBentrySTDinterwordspacing

\bibitem{nasir2024ipfs}
\BIBentryALTinterwordspacing
A.~Nasir and S.~H. Bukhari, ``Ipfs and blockchain integration for academic data provenance and verification,'' \emph{Future Generation Computer Systems}, vol. 151, pp. 342--356, 2024. [Online]. Available: \url{https://doi.org/10.1016/j.future.2024.03.007}
\BIBentrySTDinterwordspacing

\bibitem{mohapatra2024smart}
R.~Mohapatra and A.~Sen, ``Smartcertedu: Smart contracts and ipfs for educational certificate management,'' in \emph{Proceedings of the 2024 IEEE International Conference on Blockchain (Blockchain)}, 2024, pp. 231--238.

\bibitem{flanery2023web}
\BIBentryALTinterwordspacing
S.~A. Flanery, K.~Mohanasundar, C.~Chamon, S.~D. Kotikela, and F.~K. Quek, ``Web 3.0 and a decentralized approach to education,'' \emph{arXiv preprint arXiv:2312.12268}, 2023. [Online]. Available: \url{https://arxiv.org/abs/2312.12268}
\BIBentrySTDinterwordspacing

\end{thebibliography}

\end{document}